# A Policy-Aware Model for Intelligent Transportation Systems

Zacharenia Garofalaki, Dimitrios Kallergis, Georgios Katsikogiannis, Christos Douligeris
Department of Informatics, University of Piraeus, Greece
{z.garofalaki, d.kallergis, gkatsikog, cdoulig}@unipi.gr

*Abstract*—Recent advancements in the field of smart machine-to-machine (M2M) communications impose the necessity to improve the service delivery by enforcing appropriate security rules. Due to the large number of the connected devices, the criticality of the M2M applications, and the network stability weaknesses, we need to consider and analyze the security aspects and establish a flexible policy-aware architecture. This paper explores the relevant architectural challenges in this environment and proposes a Policy-Aware smart M2M Architecture (PAArc) based on ETSI's M2M communications functional architecture. We explore the policy-based management aspects to improve the security of the M2M components and services and to mitigate the security concerns that arise by evaluating an Intelligent Transportation System use case. It is shown that the policy enforcement enables enhanced security management capabilities, increased agility, and better service levels in the field of smart M2M communications.

## I. INTRODUCTION

In the last few years, smart machine-to-machine (M2M) networks have increased dramatically. The smart M2M ubiquitous computing features support low-power consumption, localized management and cope with the limited resources constraints. Concerning the network communication protocols, the smart M2M environments need to support cross-domain information exchanges among several interconnected nodes. These exchanges complicate the operations of the routing and management protocols, the communication services, and the device reachability, and raise various security and performance issues [1].

Several M2M applications (i.e. transportation, human or inventory tracking, water or energy distribution and quality monitoring, personal area networking and habitat monitoring, data center monitoring, disaster avoidance and recovery, military surveillance and industry operations, medical or healthcare monitoring, process monitoring and smart spaces) rely on the capabilities of the smart M2M communications. Policy-based management allows the creation of certain condition expressions that enable the policy enforcement on the interconnected components and resources. The considerable benefits of the policy-based management approach grow as the smart M2M communications evolve and the resources become more complex. The resources turn out to be available to the interconnected components and can be accessed using interoperable services, while the existence of a service orchestration enables the automated arrangement and management of the resources.

In this paper, we propose a *Policy-Aware M2M Architecture* (PAArc) which includes a cross-domain policy enforcement to achieve easier management and higher security of the smart M2M devices, services and communication paths. We investigate various service challenges, constraints, requirements and goals to identify the key-aspects to be considered. Additionally, we analyze an Intelligent Transportation System (ITS) use case to demonstrate the challenges of smart M2M communications.

On these grounds, this paper is based on the European Telecommunication Standards Institute (ETSI) M2M communication functional architecture [2]. Our proposal for smart M2M communications enables the enforcement of the appropriate security policies with a policy-based approach to alleviate concerns over security, and incorporates policy-based functionalities, service registry, analytics and other capabilities. The proposed PAArc secures the communication among a large population of heterogeneous smart objects, applications and services.

The remainder of the paper is organized as follows. Section 2 presents the background work, whereas the proposed architecture and the policy enforcement model is presented in Section 3. Then, Section 4 conducts a model evaluation of a use case. A discussion by contrasting the two approaches with and without PAArc is presented in section 5. Section 6 concludes the paper and presents future research directions.

## II. BACKGROUND

Several alternative architecture models akin to the OSI model have been proposed in the literature for the communication among interconnected nodes. The *ETSI M2M communication functional architecture* [2] standardizes the procedures for handling the resources and the information exchange over the M2M reference points. The ETSI M2M supports standardized security mechanisms (i.e. mutual end-point authentication, optional secure sessions, RESTful procedures). However, the ETSI M2M functional architecture does not incorporate any policy-based management to enforce the appropriate security policies in a dynamic M2M environment.

The Internet of Things – Architecture (IoT-A) [3] provides the communication needs in the IoT domain model. Although the model leverages the ISO OSI seven-layer, it is mainly focused on the interactions of the communicating systems between different stacks among the key elements, such as devices, services, applications, end-users in the device, and network domains. Cisco [4] has proposed the IoT Reference model, which consists of seven layers based on the control information flows. Still, the model does not support any security policies for monitoring and controlling different devices' communication patterns.

ETSI has established a technical body to provide the working items as a basis for liaison with other standardization organizations for the latest technology innovations in the context of M2M communications. Among these items, the ETSI M2M service requirements [5] describes the M2M communications security requirements and focuses on the M2M communications usage, the relevant technologies and methodologies. Nonetheless, the existing models lack the policy enforcement capabilities of the appropriate security controls that are necessary to establish a secure framework in the resource-constrained M2M networks.

A wide area of smart M2M implementations is the area of Intelligent Transportation Systems (ITS). The mobility of the nodes within wide areas and with high velocity combined with their resource contraints and the need for communication's availiability make the ITS valnurable to security threats [6]. The security issues of the ITS involves and affects elements (i.e. nodes, devices, services, applications, etc.) from all the architectural layers.

Regarding policy-based management models academic research, Ferraiolo et al. [7] present an innovative Policy Machine that manages access control policies independently of the hardware and software configuration. However, this work lacks the enforcement of security policies across different domains to address and resolve the latest security concerns and threats [8]. The policy-based key-elements are the following logical entities [9]:

- *Policy Enforcement Point* (PEP): performs the decision requests, receives the policy updates, translates the updates appropriately, and enforces the policy decisions.
- *Policy Decision Point* (PDP): evaluates the applicable policy against other relevant policies and attributes, and provides the decision outcome to PEP.
- *Policy Information Point* (PIP): acts as a source of attribute values to make a policy decision.
- *Policy Administration Point* (PAP): provides the authoring and maintenance of a policy or policy set(s). This includes a policy store, which is a repository for the policies.

The increasingly voluminous interconnected devices raise various scalability, interoperability and service ability concerns. To tackle this challenge, the policy tool can mitigate network failures and security attacks without necessitating the development of sophisticated protocols and mechanisms. For instance, in the case of a failure or a network topology change, it should be feasible for the policy engine to trigger an event-based policy action and to immediately and automatically reconfigure the network [10].

Barki et al. [11] provide a survey in secure M2M communications and categorize the security issues into six groups: *(a)* key management, *(b)* data-origin authentication, *(c)* entity authentication, *(d)* privacy, *(e)* data integrity and *(f)* device integrity, and they illustrate the paramount importance of the Identity Based Cryptography (IBC), and the ephemeral identity (pseudonym) in terms of privacy. Focusing on smart grids, Etigowni et al. [12] propose a logical policy enforcement by utilizing information flow tracking and logic-based context-aware policies, while SCADA systems commonly prioritize the respective security requirements such as *(a)* integrity, *(b)* confidentiality, and *(c)* trust regarding the reliability of information [13].

One of the major security concerns in policy enforcement is the detection of policy violations. Maw et al. [14] proposed a model that supports the detection of security violations by examining the respective audit record in the hosted prevention and detection mechanisms. The authors also introduced access decisions for authorization, operational and obligation policies. The authors in [15] addressed the semantic gap between the policies and the low-level mechanisms by forming a simulation apparatus with various high-level policy languages for cross-domain policy enforcement (i.e. XACML, WS-Policy). Singh et al. [16] enhanced the policy enforcement with event-based systems and addresses cross-domain policy issues, whereas Sicari et al. [17] proposed an enforcement engine which supports the management of new policies at runtime without any service disruptions. Other studies [21], [22] also demonstrate how to enforce security policies to present more secure communications models.

In the next section the Policy-Aware smart M2M Architecture is described and the policy enforcement is implemented as a tool to efficiently address the above-mentioned security concerns and management issues.

III. POLICY-AWARE SMART M2M ARCHITECTURE

*A. Architectural Model*

Based on the ETSI M2M communications functional architecture, we provide a series of policy capabilities to facilitate the enforcement of the appropriate security controls and policies for smart ubiquitous computing. To efficiently establish a security framework, we incorporate a policy-based framework. Our intention is to generalize the security policy-based enforcement at an architectural level.

In smart M2M communications, the Device domain consists mainly devices usually embedded in real-life objects to monitor or control their operations. The data gathered is transmitted to the Network Domain and, finally, to the servers in the Application domain [18]. The Network domain includes the core and access network devices and services, as well as service capabilities of the entire architecture [19]. The Application domain is a group of back-end servers that gather, store and present the sensory information via end-user applications [20]. The exchange of information between the domain's entities is facilitated by the service capabilities included in the architecture.

Due to the agile nature of the interconnected objects (i.e. nodes moving from cluster to cluster) and the cross-domain operational effects of the services, a policy-based enforcement system can simplify the dynamic service redundancy and security. This Policy-Aware Architecture (PAArc) includes a cross-domain operating policy-based service and enhances security management of the nodes in the device domain, the access and core network in the network domain, and the M2M data and various applications in the application domain.

*B. Policy enforcement*

In this section, we present the analysis of the policy enforcement throughout the architectural domains. The service provider initially publishes the services in the registry (i.e. an XML-based registry to describe, publish and utilize the services). In this vein, we describe new services to establish, or leverage existing code (e.g. JSON or binary encoding formats) for re-usable services.

The service registry maintains a set of published services with their associated service properties and the business process documents to facilitate any subsequent queries to the service registry originated by the service requester. The requester searches for a specific service in the registry and in the case of an existing and valid service forwards the request to the PEP to intercept the request into the appropriate format (i.e. using a SOAP transport message). Then, PEP invokes and binds the methods of the requested service to the service provider that provides back to PEP the outcome of the service request. Finally, PEP acts on the received policy-decision and translates the updates into the appropriate format for usage, which is then communicated to the publisher component (i.e. notice-board, APIs, text-messages and notifications).

The model allows the publishing and finding of services to resolve the service requests with policy-based criteria to enforce the appropriate security policies. More specifically, the use of the service data enables the offloading of the metadata and of the service attributes into the service data repository providing increased flexibility of specifying many transactions with consistency, re-usability, and faster implementation in generic processing.

Upon receiving a request by the service-provider, PDP evaluates the service-request based on the managed policies already defined by the policy manager. PAP feeds the policy-store repository with the policies. The relevant policies to the specific service-request are retrieved and consulted by PDP to reach a decision. If there is no blocking or negative policy decision, PDP retrieves the necessary additional service data from PIP, which offers the interface to collect the appropriate service data related to the service-request.

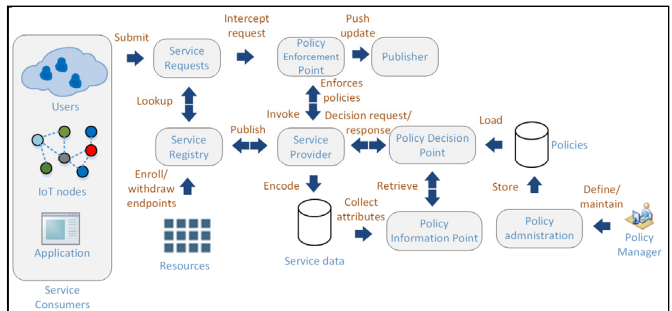

Figure 1. Policy enforcement methods

The policy-decision, supported by the service attributes, is forwarded to the service-provider to resolve the request. The detailed structure of the policy enforcement methods and of the corresponding abstract interactions is shown in Figure 1, which illustrates the entities' interactions and the detailed messages exchange in a time-order manner.

The policy enforcement is facilitated by the use of services, such as a Trusted Certificate Authority (TCA or CA) for issuing digital certificates, a Registration Authority (RA) for verifying the endpoint request for digital identity, a Validation Authority (VA) to verify the validity of the digital certificate with the appropriate technical means.

IV. USE CASE AND ARCHITECTURAL MODEL EVALUATION

In this section, through a use-case we describe how we can integrate the policy enforcement on the architecture. We scrutinize the architecture for a use case and we elaborate on the architecture's security requirements and concerns for secure smart M2M communications.

In this context, we analyze the implementation of the security policies and workflows for the intelligent Bus on Campus (iBuC) [23], which has been proposed to offer a state-of-the-art transportation service within a university campus utilizing multiple AVs, smart devices, and the existing wireless infrastructure and mobile services. The iBuC service requests are forwarded via a booking application to the control unit (CU) which accordingly calculates both the AV's and the passenger's estimated time of arrival (ETA) to the bus stop and elects an eligible vehicle to operate an itinerary for academic passengers. The CU also collects and analyses various datasets regarding the vehicles' and passengers' status by using wireless sensor devices, GPS and Wi-Fi communication capabilities.

To evaluate the architectural model discussed previously, we contrast the iBuC operational scenario with an extended scenario where iBuC integrates the policy enforcement of security policies. In the following section, case A presents the workflows based on the original iBuC work and case B shows the messages exchanged with the PPArc architecture.

*A. Enroll/withdraw resources*

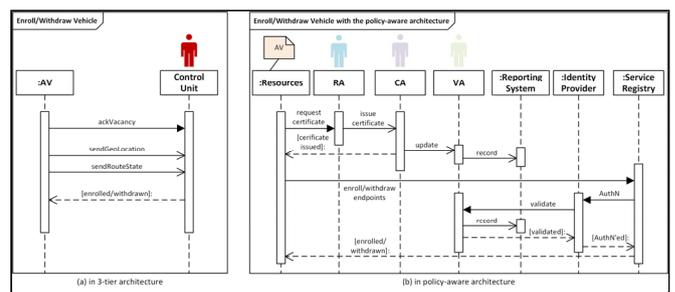

Figure 2. Enroll or withdraw an AV

Figure 2 depicts how an AV can be enrolled to or withdrawn from the transportation service with reliable registration activities. In case A, the AV provides real-time data and the state to the Control Unit (CU). Along with the location, route-paths, service-bulletins, stop-list and the current route state (e.g. running state, idle state, near finish state) that can be collected by the M2M devices through the M2M network (i.e. wireless sensor network), the CU records entering-AVs (i.e. operate the next itinerary) or leaving-AVs (i.e. to be substituted by another one) to leverage efficient AV services.

In case B and in the context of the policy-aware architecture, multiple entities enable the secure identification of the AVs and enhance the security based on digital identity assertions. The AVs utilize digital certificates to be authenticated and securely registered into the service. The message exchanges are securely protected with the encryption security mechanisms, and the solution enables to securely store, retrieve, analyze, and integrate data.

## V. Discussion

Considering the recent rapid advances and the respective security challenges, the reasoning in proposing a policy-aware architecture (PAArc) is to introduce policy-based management to facilitate the enforcement of the security policies to comply with the security objectives. One of the major challenges for smart M2M communications is to support a large device volume with various heterogeneous characteristics and different mobility profiles by meeting the resource constraints requirements. For instance, energy efficiency is a main design objective, and, thus, the smart M2M devices should operate efficiently with low latency, high resilience and high service reliability considering the critical nature of several smart M2M applications (e.g. in healthcare, critical infrastructures and transportation). The various security concerns mentioned above should be addressed, and we need to implement secure M2M services by design advocated with built-in security in the smart M2M nodes and the conventional application security solutions. In more detail, we need a policy enforcement to apply the appropriate security policies and to ensure data privacy and effective countermeasures in a dynamic environment.

Our contribution in PAArc is that we manage to enforce the proper security policies in the smart M2M communications services to strengthen and improve a robust security management. Among several other security solutions, the access to sensitive data, the data confidentiality and integrity, and the entity authentication can be imposed with policy-based methods. In the context of distributed, multi-domain and often unattended M2M networks, we advocate that hierarchical policies offer an innovative security policy approach in policy composition, enforcement, validation, and conflict resolution. In these cases, various complicated issues also arise such as the semantic interoperability, the definition of concrete policies, multi-domain policy consistency, security policy refinement, and policy and completeness, which are not subject of this work.

The proposed PAArc offers several enhancements in the service composition and improved information flows. First, the model-driven development facilitates the business intelligence and optimizes performance functionality. With the policy-aware model, the certificate issuance policies can be used for a qualified subordination between different PKI hierarchies (i.e. recognize certificates by another CA that meet the certificate issuance requirements). Along with the ability to expose the functionalities, this model provides additional levels of security and robustness. The audit records are kept in the reporting system, which allows the construction of a comprehensive range of traceability and serviceability queries.

In terms of manageability, the prototype of PAArc is re-usable and enables the enforcement of security policies and the development of new functionalities easily and rapidly. Various requirements and service levels can be encoded with clear service descriptions for service reusability. PAArc optimizes the development efforts of different use case configurations (i.e. AV initial setup) and complex event processing to adapt to various factors and environments (i.e. moving to another place with diverging climatic and environmental conditions). Besides, PAArc supports message transformation with data translation from the canonical representation into the technical form, which simplifies the software structure needed for the implementation of the proposed model. Not only the combination of various service metadata is supported, but we also manage to increase the flexibility and integration capabilities with other service modules with data sharing.

In terms of security policy enforcement, enhanced message security with respect to data confidentiality and integrity is also achieved. The data-origin authentication guarantees that the message has a distinct origin entity whilst entity authentication facilitates the communication entity to prove its legitimate identity. The smart M2M communications rarely rely on static node infrastructures and, thus, these communications require secure transport layer protocols to establish a reliable and secure end-to-end communication path. These enhancements and improvements along with the application security and management policies show the added-value of PAArc.

## VI. Conclusions

The high number of interconnected heterogeneous devices raises several security and scalability issues, as we need to employ the appropriate security countermeasures and policies to achieve the appropriate levels of secure services. In this paper, we described the ETSI functional M2M communications 3-tier architecture with the addition of the policy enforcement across the domains. The proposed Policy-Aware Architecture (PAArc) facilitates the evolution of the appropriate security policies, and provides a secure interoperability between the device, network and application domains, and efficient monitoring mechanisms. Addressing and managing efficiently the security concerns for each domain of the proposed architecture with a policy tool and by incorporating the smart M2M service requirements, we manage to improve the security service levels of M2M communications.

The use of policy-based management has several advantages in implementing adaptive smart M2M communications, where the network topology can dynamically change in response to the mobility of the interconnected objects. A policy-based management approach ensures the enforcement of the security policies to mitigate the security issues, the risks and improves the security controls and data privacy. The policy enforcement engine has access to the security policies and the additional information, and then applies the respective policies at fixed points in the device, network and application domains. Therefore, the establishment and the enforcement of clear and effective security and privacy policies are key issues to improve the offered service levels in smart M2M communications.

Policy-based management fosters the enforcement of service-oriented cognitive technologies in smart M2M communications. In this regard, our future work will focus on the integration capabilities of the dynamic service registry, service location discovery, and the policy manager as the key functionalities of an integrated and secure policy-based architecture in smart M2M communications.

ACKNOWLEDGMENT

The research presented in this paper has been partly supported by the Erasmus+ project Open Models Initiantive – OMI, Key Activity 2. The European Commission support for the production of this publication does not constitute an endorsement of the contents which reflects the views only of the authors, and the Commission cannot be held responsible for any use which may be made of the information contained therein.